\documentclass[preprint,11pt]{elsarticle}

\usepackage{microtype}
\usepackage{subcaption}
\usepackage{float}
\usepackage{graphicx}
\usepackage{multirow}
\usepackage{booktabs}
\usepackage[table]{xcolor}

\usepackage{amssymb}
\usepackage{amsmath}

\journal{ } 
\date{ }

\begin{document}
	
	\begin{frontmatter}
		
		\author{Ju-Hong Lee\fnref{label1}}
		\ead{juhong@inha.ac.kr}
		\author{Bayartsetseg Kalina\fnref{label1}}
		\ead{kb0422.bk@gmail.com}
		\author{KwangTek Na\fnref{label1}}
		\ead{kwangteakna@gmail.com}
		\affiliation[label1]{organization={Inha University},
			city={Incheon},
		 	postcode={22212},
		 	country={South Korea}}
		
		\title{Market-adaptive Ratio for Portfolio Management}
		
		
		
		
		\begin{abstract}
			Traditional risk-adjusted returns, such as the Treynor, Sharpe, Sortino, and Information ratios, have been pivotal in portfolio asset allocation, focusing on minimizing risk while maximizing profit. Nevertheless, these metrics often fail to account for the distinct characteristics of bull and bear markets, leading to sub-optimal investment decisions. This paper introduces a novel approach called the Market-adaptive Ratio, which was designed to adjust risk preferences dynamically in response to market conditions. By integrating the $\rho$ parameter, which differentiates between bull and bear markets, this new ratio enables a more adaptive portfolio management strategy. The $\rho$ parameter is derived from historical data and implemented within a reinforcement learning framework, allowing the method to learn and optimize portfolio allocations based on prevailing market trends. Empirical analysis showed that the Market-adaptive Ratio outperformed the Sharpe Ratio by providing more robust risk-adjusted returns tailored to the specific market environment. This advance enhances portfolio performance by aligning investment strategies with the inherent dynamics of bull and bear markets, optimizing risk and return outcomes. 
		\end{abstract}
		
		
		\end{frontmatter}
		
		\section{Introduction}
		
		Portfolio asset allocation is a crucial in determining how capital is distributed among various assets to achieve investment objectives, primarily minimizing risk and maximizing profit. Traditional risk-adjusted returns, such as the Treynor, Sharpe, Sortino, and Information ratios have been used extensively to evaluate portfolio performance relative to risk. These measures help mitigate the tendency of investors to focus solely on returns without considering the associated risks. Nevertheless, they often overlook an essential aspect of financial markets: the distinction between bear and bull markets. 
		
		In financial terminology, a bull market is characterized by rising asset prices, encouraging investors to pursue higher returns. In contrast, a bear market signifies declining asset prices, prompting investors to minimize risks and preserve capital. The conventional one-size-fits-all approach of traditional risk-adjusted returns does not capture these market types, leaving investors with an incomplete picture of portfolio performance. This paper introduces a novel metric, the Market-adaptive Ratio, to address this gap by incorporating market-specific risk preferences into portfolio management.

		\section{Related work}
		
		\subsection{Traditional Risk-Adjusted Returns}
		The Treynor \cite{Treynor65}, Sharpe \cite{Sharpe66}, Sortino \cite{Sortino94}, and Information ratios have been the cornerstones of portfolio performance evaluations. These metrics assess returns relative to various risk measures but fail to distinguish between bull and bear market conditions.
		
		\subsubsection{Sharpe Ratio}
		The Sharpe Ratio \cite{Sharpe66} is used widely to evaluate the risk-adjusted return of a portfolio. One significant advantage of the Sharpe Ratio is that it does not require a proxy for the market portfolio. The Sharpe Ratio relates the expected excess returns of a portfolio to its risk, which is represented by the standard deviation of the excess return:
		\begin{equation*}
		\text{Sharpe Ratio}=\frac{\mu_t-R_f}{\sigma_t}
		\end{equation*}
		where $\mu_t$ is the expected excess return of a portfolio at time $t$; $R_f$ is the risk-free rate of return, and $\sigma_t$ is the standard deviation of the excess returns of a portfolio at time $t$, which measures the portfolio risk. 
		
		\subsubsection{Treynor Ratio}
		The Treynor Ratio uses beta (systematic risk measure) instead of the standard deviation to assess performance. 
		\begin{equation*}
		\text{Treynor Ratio}=\frac{\mu_t-R_f}{\beta_t}
		\end{equation*}
		where $\mu_t$ is the expected excess return of a portfolio at time $t$; $R_f$ is the risk-free rate of return, and $\beta_t$ is the beta of a portfolio at time $t$.
		
		The Treynor Ratio requires a benchmark market index, which may not be readily available or relevant for all portfolios. 
		
		\subsubsection{Sortino Ratio}
		The Sortino Ratio focuses on downside risk using the standard deviation of negative returns. 
		\begin{equation*}
		\text{Sortino Ratio}=\frac{\mu_t-R_f}{\bar{\sigma}_t}
		\end{equation*}
		where $\mu_t$ is the expected excess return of a portfolio at time $t$; $R_f$ is the risk-free rate of return, and $\bar{\sigma}_t$ is the downside deviation of the excess returns of a portfolio at time $t$.
		
		While this is useful for investors more concerned with losses, it requires a subjective determination of a "negative" return. In addition, it does not penalize portfolios for taking on excessive risk in the pursuit of high returns, potentially encouraging riskier behavior. 
		
		\subsubsection{Information Ratio}
		The Information Ratio compares the excess return of a portfolio to its tracking error (standard deviation of excess returns relative to a benchmark). 
		\begin{equation*}
		\text{Information Ratio}=\frac{\mu_t-R_b}{\sigma_t(\mu_t-R_b)}
		\end{equation*}
		where $\mu_t$ is the expected excess return of a portfolio at time $t$; $R_b$ is the return of a chosen benchmark, and $\sigma_t(\mu_t-R_b)$ is the active risk of the portfolio at time $t$.
		
		This ratio requires a relevant benchmark and is mainly useful for active managers evaluated against a specific index. The Information Ratio may not be as applicable for diverse or unconventional portfolios without a clear benchmark.
		
		\subsection{Portfolio Asset Allocation}
		Markowitz's Modern Portfolio Theory (MPT) \cite{Markowitz52} and risk budgeting \cite{Roncalli14, Richard19} provide foundational frameworks for portfolio construction. These methods use historical data to estimate expected returns, variances, and covariances of asset returns. This method involves calculating the optimal portfolio weights that maximize the Sharpe Ratio based on these historical estimates. These weights are then applied to future data, assuming that past statistical relationships remain stable. This static approach, while simple and well-understood, lacks the flexibility to adapt to changing market conditions, and the performance can suffer if the assumption of market stationarity does not hold. 
		
		Reinforcement learning (RL)-based portfolio asset allocation methods, such as recurrent reinforcement learning (RRL) \cite{Moody98, LinLi21}, have introduced more adaptive approaches to portfolio management. They commonly use the Sharpe Ratio \cite{Sharpe66} as the reward function. RRL uses a dynamic and adaptive learning process. During the training phase, RRL methods use historical data to learn optimal portfolio allocation strategies, adjusting continuously based on the observed returns and risks. This is achieved by maximizing the Sharpe Ratio, through reinforcement learning algorithms. When applied to test data, the trained RRL method continues to adapt, updating its strategy in response to new market information. This makes RRL more robust in non-stationary markets because it can adjust to changing conditions. On the other hand, using the Sharpe Ratio as a reward function in RRL presents significant disadvantages. Primarily, the inability of the Sharpe Ratio to distinguish between bull and bear markets results in a uniform risk assessment that does not adapt to market conditions, potentially leading to suboptimal portfolio strategies during periods of market volatility. This metric treats all risks equally, ignoring the differing risk preferences investors might have in varying market environments, such as taking on more risk in bull markets to maximize returns and minimizing risk in bear markets to preserve capital.

		\section{Limitation of Traditional Risk-Adjusted Returns}
		Traditional risk-adjusted returns do not account for market phases, leading to sub-optimal investment decisions. For example, the Sharpe Ratio may not adequately guide portfolio selection when multiple portfolios exhibit the same ratio, particularly under varying market conditions. In bull markets, investors might prefer higher-risk portfolios for potentially higher returns, whereas the focus shifts to capital preservation in bear markets. 
		
		\begin{figure}[H]
		\begin{center}
			\centerline{\includegraphics[width=0.95\columnwidth]{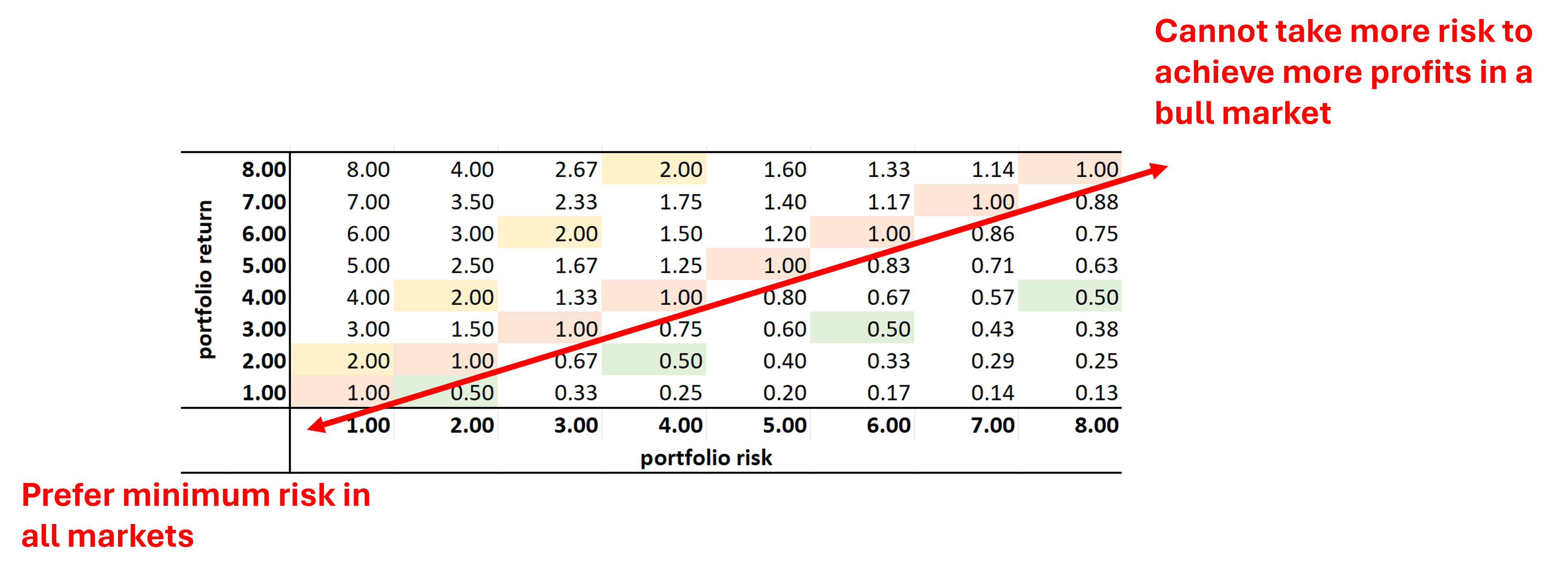}}
			\caption{Sharpe Ratio}
			\label{fig_1}
		\end{center}
		\end{figure}
		
		The presence of several portfolios with a Sharpe Ratio of 1.00 indicates that they have all achieved similar levels of risk-adjusted returns, as shown in Figure 1. Nevertheless, while the Sharpe Ratio offers insights into the risk-return tradeoff, it does not definitively determine which portfolio is superior, particularly when multiple portfolios share the same value. The experimental results showed that the portfolio with the lowest risk (a Sharpe Ratio with $\sigma_t=1$ in Figure 1) is selected. This decision is based on the premise that lower risk typically aligns with the goal of capital preservation, which is crucial for investors, especially under uncertain market conditions. 
		
		Selecting the lowest-risk portfolio does not guarantee optimal performance in terms of maximizing returns or minimizing risk across different market types. Market dynamics, such as those observed in bull and bear markets, introduce additional complexities that the Sharpe Ratio cannot be fully captured. During bull markets, characterized by rising asset prices, the emphasis may be on capitalizing on opportunities for higher returns, even if it entails more risk. For example, a Sharpe Ratio with $\sigma_t=8$ during a bull market would be preferred over a Sharpe Ratio with $\sigma_t=1$, as shown in Figure 1. 
		
		In contrast, minimizing risk becomes paramount in bear markets where asset prices are declining, and investors prioritize capital preservation. For example, a Sharpe Ratio with $\sigma_t=1$ during a bear market would be preferred over a Sharpe Ratio with $\sigma_t=8$. Therefore, while the Sharpe Ratio provides valuable insights into risk-adjusted performance, it is essential to consider additional factors, such as market conditions when making portfolio allocation decisions.
		
		\section{Market-adaptive Ratio}
		The limitation of the Sharpe Ratio was addressed in this study by introducing a novel reward function for RL-based portfolio asset allocation called the Market-adaptive Ratio. This ratio incorporates the $\rho$ parameter to adjust risk preferences dynamically according to market conditions. The $\rho$ parameter is defined as:
		\begin{equation}
		\rho_t=\frac{2}{1+e^{-\alpha R_t}}
		\end{equation}
		where $R_t$ is the return at time $t$, and $\alpha$ is a hyperparameter. The Market-adaptive Ratio is calculated as follows:
		\begin{equation}
		m_t=\frac{sgn(\mu_t-R_f)|\mu_t-R_f|^{\rho_t}}{(\sigma_t)^{1/\rho_t}}
		\end{equation}
		where $\mu_t$ is the portfolio return; $R_f$ is the risk-free rate of return; $\sigma_t$ is the portfolio risk; $\rho_t$ adapts to market conditions, allowing risk-taking in bull markets and risk minimization in bear markets. 
		
		The Market-adaptive Ratio is defined as an asymmetric risk-adjusted return that incorporates the $\rho$ parameter. This parameter represents the distinct attributes of bear and bull markets, enabling a dynamic adjustment of the risk preferences. During bull markets, the Market-adaptive Ratio allows for taking risks to capitalize on market upswings, whereas it aims to minimize risk to protect portfolios from losses during bear markets. By integrating the $\rho$ parameter, the Market-adaptive Ratio offers a tailored approach to portfolio asset allocation that adapts to changing market conditions. 
		
		The Market-adaptive Ratio in RL-based portfolio asset allocation was implemented using historical data to estimate the $\rho$ parameter. Subsequently, the Market-adaptive Ratio was incorporated as the reward function in the RL framework, allowing optimal portfolio allocation strategies to be learned by the agent, which adapts to prevailing market conditions. The effectiveness of the Market-adaptive Ratio was validated through empirical analysis, and its performance was compared against the RRL with Sharpe Ratio.
		
		\begin{figure}[H]
		\begin{center}
			\centerline{\includegraphics[width=0.9\columnwidth]{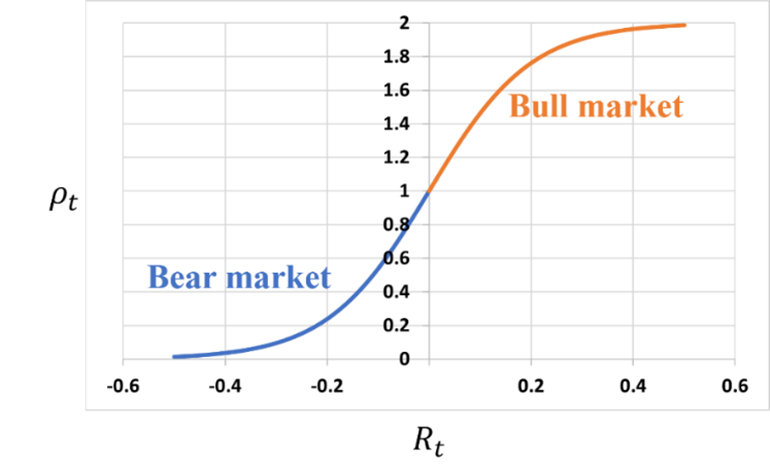}}
			\caption{Domain of the $\rho$ parameter}
			\label{fig_2}
		\end{center}
		\end{figure} 
		
		The key idea behind applying the $\rho$ parameter is to differentiate between the goals of different market conditions-specifically, taking on more risk during bull markets and minimizing risk during bear markets. The domain of the $\rho$ parameter (Figure 2) is set between 0 and 2, allowing it to adapt to different market conditions. A $\rho$ parameter is close to 0 indicates a bear market. In this scenario, the model prioritizes minimizing risk. A $\rho$ parameter is close to 2 indicates a bull market. In this scenario, the model permits taking on more risk to maximize returns.
		
		\begin{figure}[H]
		\begin{center}
			\centerline{\includegraphics[width=0.95\columnwidth]{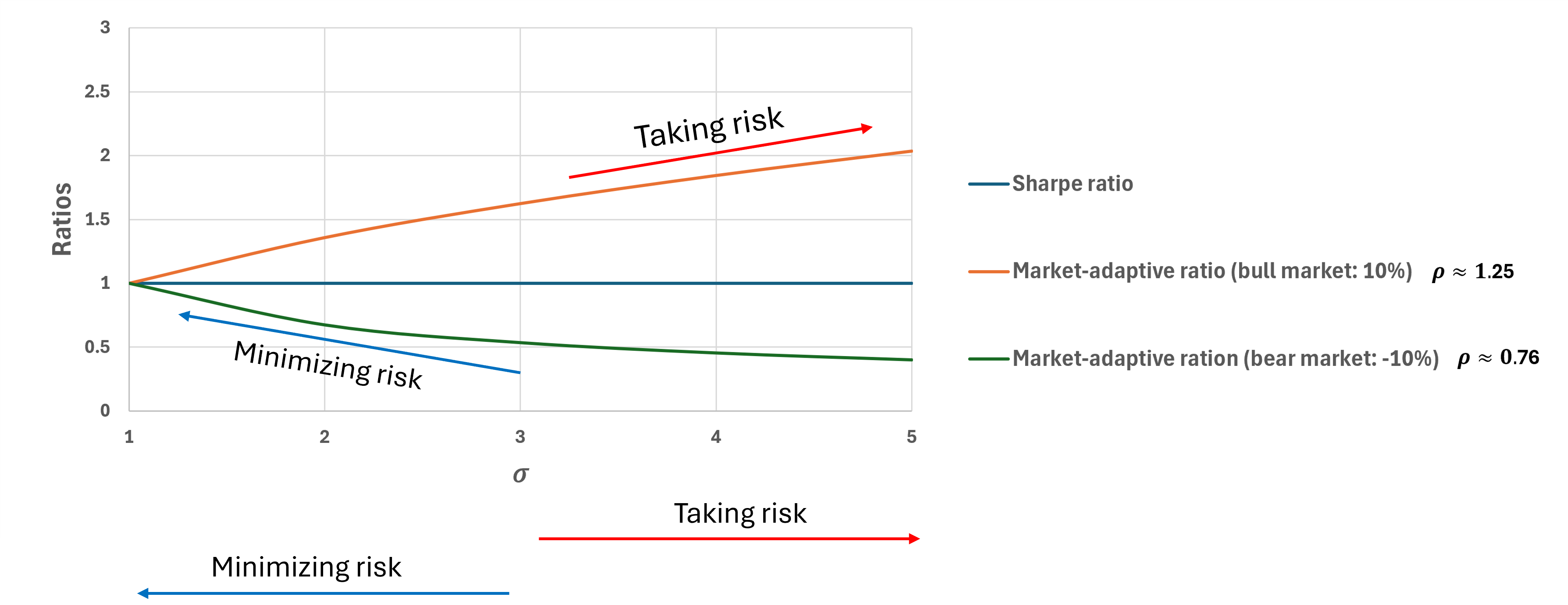}}
			\caption{Comparison of the Sharpe Ratio and Market-adaptive Ratio}
			\label{fig_3}
		\end{center}
		\end{figure} 
		
		Figure 3 compares the Sharpe and Market-adaptive ratios (where the Sharpe Ratio equals to 1.00). The Sharpe Ratio was 1.00 in all risk levels. In contrast, the Market-adaptive Ratio increased in a bull market when $\rho\in(1,2)$, enabling risk-taking. By contrast, in a bear market, it decreased when $\rho\in(0,1)$, minimizing risk. Hence, the domain of the $\rho$ parameter, ranging from 0 and 2, allows for a flexible, adaptive approach to portfolio asset allocation that considers the differing objectives of bull and bear markets.  	
		
		\begin{figure}[H]
			\label{fig_4}
			\centering
			\begin{subfigure}[b]{0.95\columnwidth}
				\centering
				\includegraphics[width=\columnwidth]{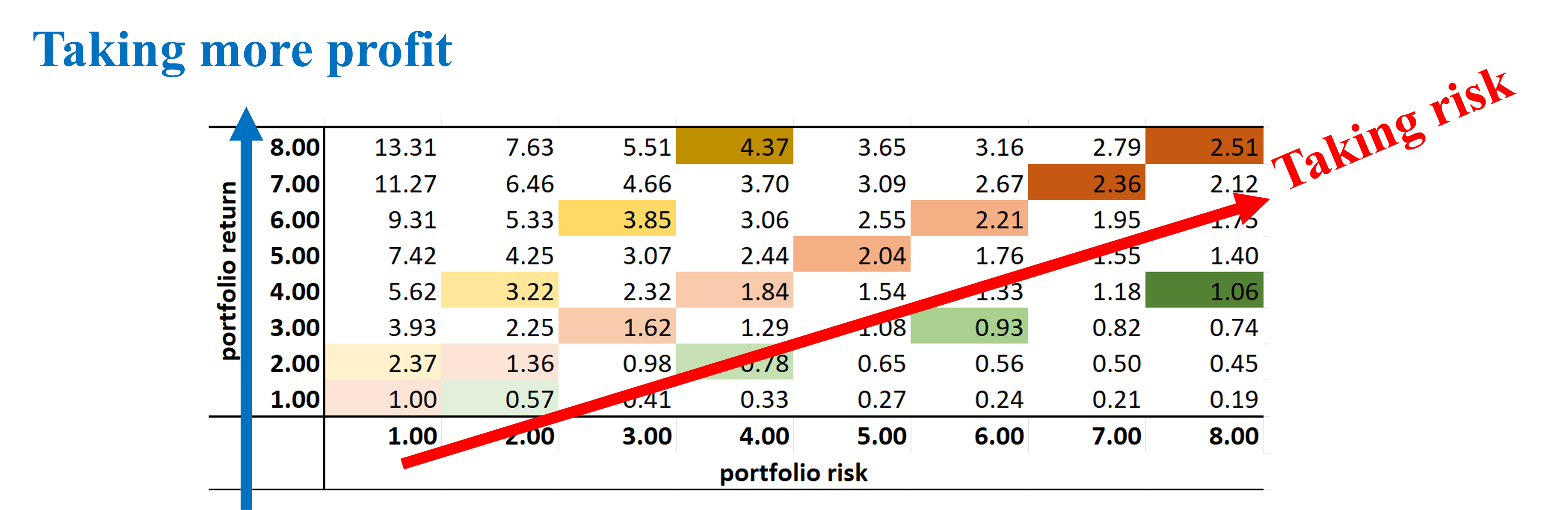}
				\caption{$R_t=10\%$, $R_f=0$, and $\alpha=5$}
				\label{fig_4a}
			\end{subfigure}
			\begin{subfigure}[b]{0.95\columnwidth}
				\centering
				\includegraphics[width=\columnwidth]{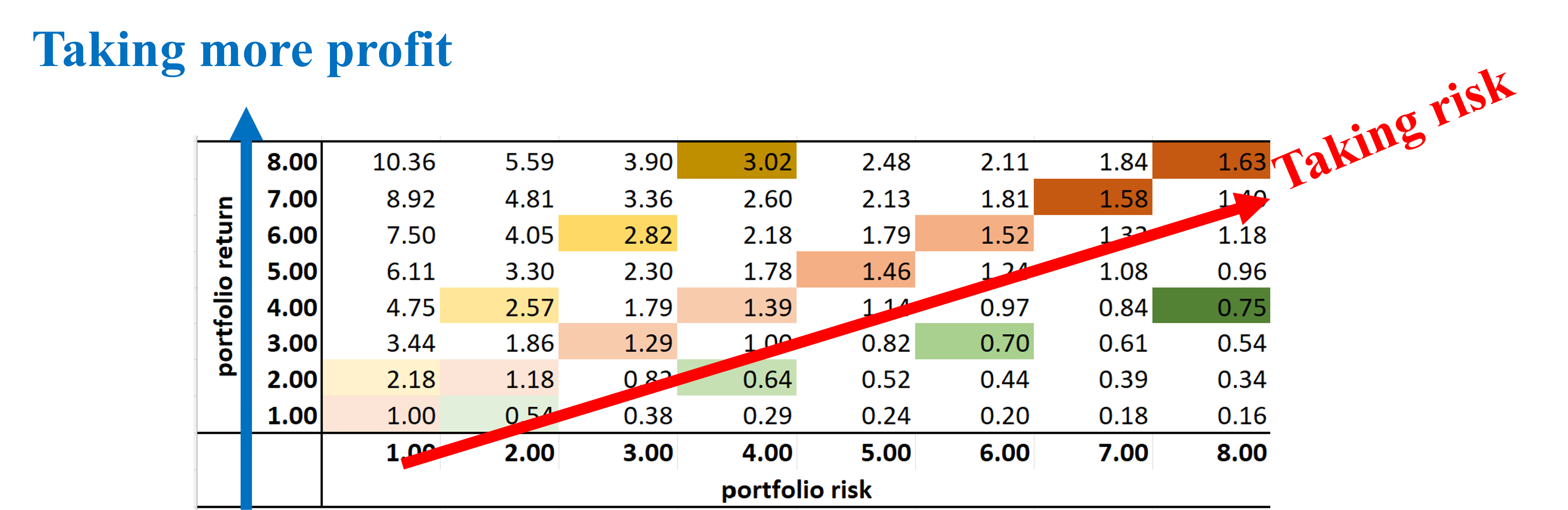}
				\caption{$R_t=5\%$, $R_f=0$, and $\alpha=5$}
				\label{fig_4b}
			\end{subfigure}
			\begin{subfigure}[b]{0.95\columnwidth}
				\centering
				\includegraphics[width=\textwidth]{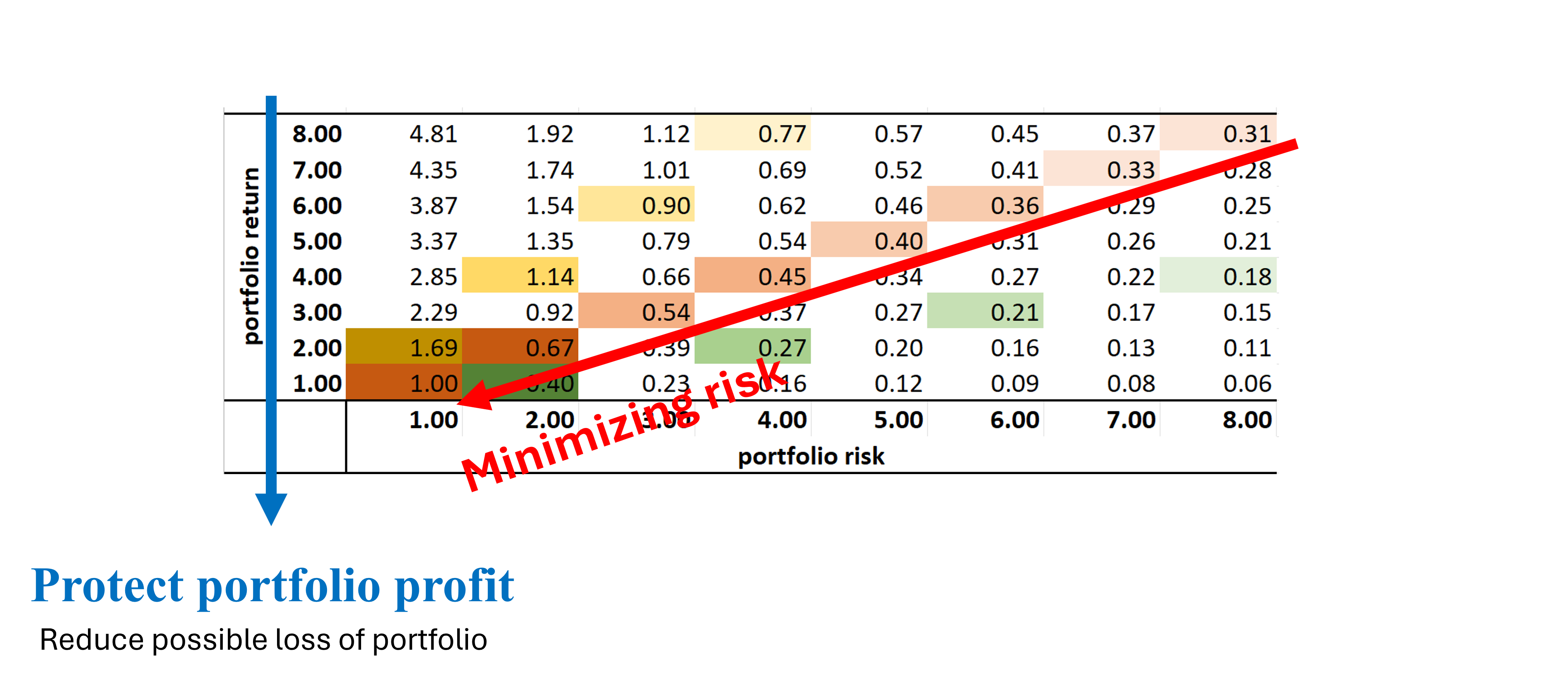}
				\caption{$R_t=-10\%$, $R_f=0$, and $\alpha=5$}
				\label{fig_4c}
			\end{subfigure}
			\begin{subfigure}[b]{0.95\columnwidth}
				\centering
				\includegraphics[width=\textwidth]{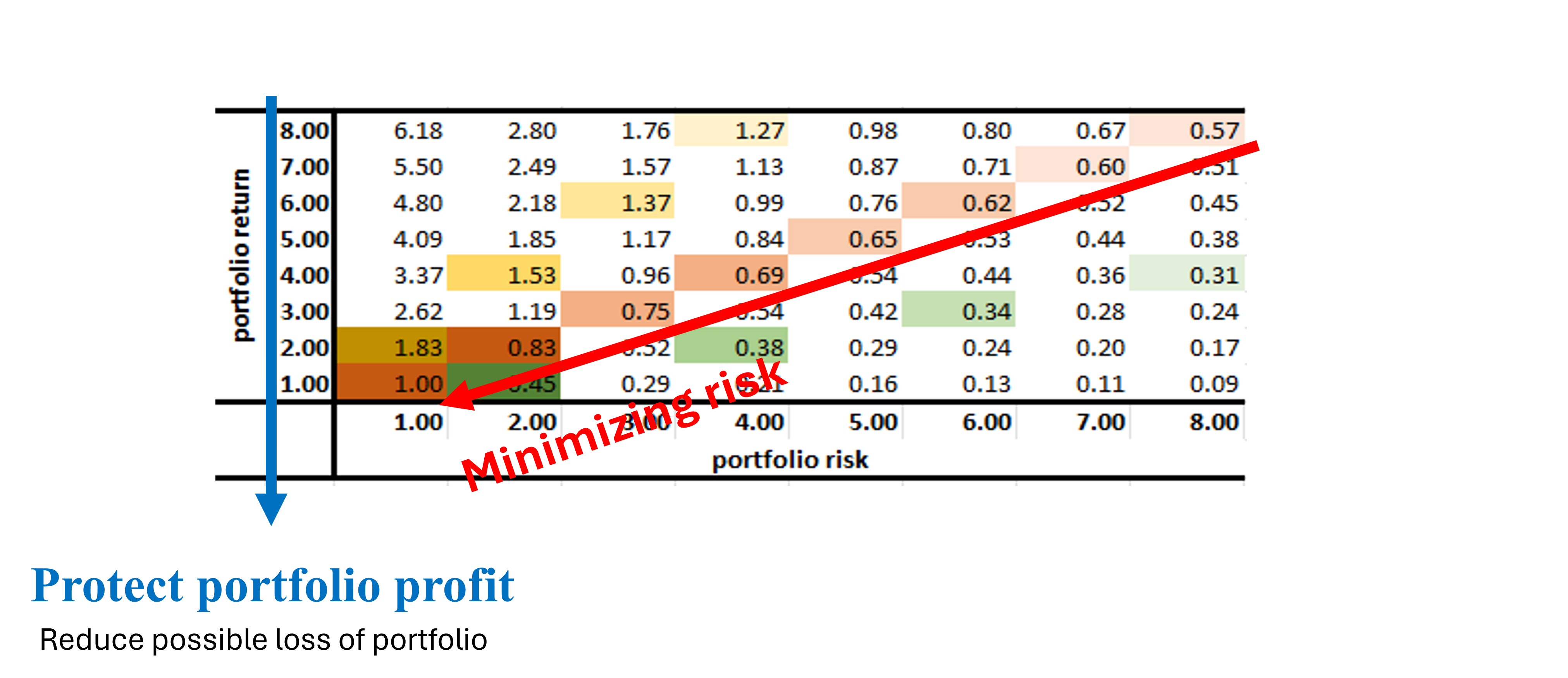}
				\caption{$R_t=-5\%$, $R_f=0$, and $\alpha=5$}
				\label{fig_4d}
			\end{subfigure}
			\caption{Examples of the Market-adaptive Ratio}
		\end{figure}
		
		The Market-adaptive Ratio represents a significant advance in RL-based portfolio asset allocation by incorporating market-type-specific risk preferences. The Market-adaptive Ratio enables investors to achieve more robust and adaptive portfolio allocation strategies by dynamically adjusting risk preferences based on bear and bull market conditions. The values differ based on the market type, reflecting the distinct characteristics of bull and bear markets (Figure 4). During bull markets (Figures 4a and 4b), characterized by rising asset prices, the Market-adaptive Ratio permits taking risks, allowing the portfolios to allocate more resources to riskier assets. For example, a Market-adaptive Ratio of 2.51 with $\sigma_t=8$ is better than a Market-adaptive Ratio of 1.00 with $\sigma_t=1$, as shown in Figure 4a. On the other hand, during bear markets (Figures 4c and 4d), marked by declining asset prices, the primary objective is to minimize risk. For example, a Market-adaptive Ratio of 1.00 with $\sigma_t=1$ is better than a Market-adaptive Ratio of 0.31 with $\sigma_t=8$ (Figure 4c). Under these market conditions, the Market-adaptive Ratio adjusts accordingly, aiming to minimize risk by reallocating assets to safer investments with lower volatility and downside protection. By recognizing the distinct attributes of bull and bear markets and adapting portfolio allocations accordingly, the Market-adaptive Ratio seeks to optimize risk-adjusted returns and enhance portfolio performance across different market environments.
		
		\section{Experiment}
		\subsection{Dataset}
		This study evaluated the BND\&SP500 portfolio (Vanguard Total Bond Market Index Fund and S\&P500 Index) and the BSV\&DAX portfolio (Vanguard Short-Term Bond Market Index Fund and DAX Index)\footnote{https://finance.yahoo.com/}. Daily data from January 2010 to December 2022 were used, with methods pretrained on data from January 2010 to December 2014 and tested on data from January 2015 to December 2022. A rolling window approach was implemented to retrain the methods annually. The portfolios were rebalanced monthly.
		
		\subsection{Method and Benchmark Comparison}
		The proposed RRL with Market-adaptive Ratio was compared against equally weighted portfolios, tangency portfolios, risk budgeting, and RRL with the Sharpe Ratio. 
		\begin{itemize}
		\item \textbf{Equally Weighted Portfolio}: This portfolio allocates equal weights to both asset types in the portfolio.
		\item \textbf{Tangency Portfolio}: Derived from Modern Portfolio Theory (MPT), this portfolio maximizes the Sharpe Ratio.
		\item \textbf{Risk Budgeting}: This method allocates risk among different assets based on predefined risk constraints to control the overall risk of the portfolio.
		\item \textbf{RRL with Sharpe Ratio}: This is the traditional RRL approach using the Sharpe Ratio as the reward function.
		\end{itemize}
		Performance metrics during the test period include the following: Profit (higher is better), represented by the expected portfolio return; Risk (lower is better), represented by the standard deviation of the portfolio return; the Sharpe Ratio (higher is better), which evaluates the risk-adjusted return by comparing the expected portfolio return to its standard deviation.
		
		\subsection{Experimental Results}
		Tables \ref{table_1} and \ref{table_2} list the experimental results for the BND\&SP500 and BSV\&DAX portfolios, respectively.  
		
		\begin{table}[H]
		\caption{Results of the BND\&SP500 portfolio \label{table_1}}
		\resizebox{\columnwidth}{!}{
			\begin{tabular}{llll}
				\toprule
				\multirow{2}{*}{Model} & Profit & Risk & Sharpe Ratio  \\
				& (Higher the Better) & (Lower the Better) & (Higher the Better) \\
				\midrule
				Equally Weighted				& 0.3833 			& 0.2779 			& 1.3793 \\
				Tangency portfolio 				& \textbf{0.5587}	& 0.4183 			& 1.3356 \\
				Risk Budgeting 					& 0.2303 			& \textbf{0.2126}	& 1.0835 \\
				RRL with Sharpe Ratio			& 0.2866 			& 0.2483 			& 1.1540 \\
				RRL with Market-adaptive Ratio	& 0.3218 			& 0.2248 			& \textbf{1.4318} \\
				\bottomrule
		\end{tabular}}
		\end{table}
		
		\begin{table}[H]
		\caption{Results of the BSV\&DAX portfolio \label{table_2}}
		\resizebox{\columnwidth}{!}{
			\begin{tabular}{llll}
				\toprule
				\multirow{2}{*}{Model} & Profit & Risk & Sharpe Ratio  \\
				& (Higher the Better) & (Lower the Better) & (Higher the Better) \\
				\midrule
				Equally Weighted				& 0.1948 			& 0.3070 			& 0.6346 \\
				Tangency portfolio 				& \textbf{0.3822}	& 0.5053 			& 0.7564 \\
				Risk Budgeting 					& 0.0782 			& \textbf{0.0947}	& 0.8264 \\
				RRL with Sharpe Ratio			& 0.0421 			& 0.1884 			& 0.2235 \\
				RRL with Market-adaptive Ratio	& 0.1524 			& 0.1687 			& \textbf{0.9038} \\
				\bottomrule
		\end{tabular}}
		\end{table}
		
		The empirical results showed that the RRL with Market-adaptive Ratio consistently outperformed the benchmark methods in both portfolios, achieving higher Sharpe Ratios and demonstrating superior risk-adjusted performance.
		
		\section{Conclusion}
		This paper presented the Market-adaptive Ratio, a novel metric incorporating market-specific risk preferences to enhance portfolio performance. By dynamically adjusting risk preferences based on market conditions, the Market-adaptive Ratio provides a more responsive and adaptive approach to RL-based portfolio asset allocation. The empirical results validated its effectiveness, highlighting its potential to optimize risk-adjusted returns across different market environments. Consequently, the Sharpe Ratio of the proposed method during the test period was the highest across both portfolios, indicating superior risk-adjusted performance.
		
		In conclusion, the Market-adaptive Ratio represents a significant advance in RL-based portfolio asset allocation, offering a robust tool for investors seeking to navigate the complexities of dynamic financial markets. The ability of the Market-adaptive Ratio to enhance risk-adjusted returns by accounting for market conditions marks a crucial step forward in developing adaptive and responsive investment strategies.
		
\bibliographystyle{plain} 
\bibliography{example_paper}
	
\end{document}